\documentclass[12pt]{iopart}
\usepackage{graphicx}
\begin{document}
\def \lt{\!\!<\!}

\title[Deceleration without Dark Matter]{Deceleration without Dark Matter}

\author{J C Jackson and Marina Dodgson\dag
\footnote[2]{To whom correspondence should be addressed
(john.jackson@unn.ac.uk)}}

\address{\dag Department of Mathematics and Statistics,
University of Northumbria at Newcastle, Ellison Place, Newcastle NE1 8ST, UK}

\begin{abstract}

\noindent In homogeneous isotropic cosmological models the angular size of a standard measuring rod 
changes with redshift z in a manner which depends upon the parameters of the model. It has been 
argued that as a population ultra­compact (milliarcsecond) radio-sources measured by very long-baseline 
interferometry (VLBI) do not evolve with cosmic epoch, and thus comprise a set of standard 
objects, at least in a statistical sense. Here we examine the angular-size/redshift relation for 256 
ultra-compact sources with z in the range 0.5 to 3.8, for cosmological models with two degrees of 
freedom($\Omega_0$ and $\Lambda_0$). The canonical inflationary CDM model ($\Omega_0 = 1$, $\Lambda_0 = 0$) 
appears to be ruled out by the observed relationship, whereas low-density models with a cosmological
constant of either sign are favoured.

\medbreak\noindent 
Key words: cosmology -- observations -- theory -- dark matter. 

\end{abstract}

\section{Introduction}

With regard to our beliefs about the density of the Universe, the current situation might be de­ 
scribed as a binary one; the canonical inflationary spatially flat CDM model, with $\Omega_0 = 1$ 
and $\Lambda_0 = 0$, has great theoretical appeal, but little direct observational support,
whereas direct observational evidence indicates a value no greater than $\Omega_0\sim 0.2$
(that is if we include a generous estimate of the dark matter indicated by the dynamics
of clusters of galaxies), which figure has no theoretical underpinning. Doubts about the
viability of the canonical model have been reinforced by a resurgence of the cosmological
age problem (Bolte \& Hogan 1995), brought about by recent determinations of Hubble's constant
using extra­galactic Cepheids (Pierce et al. 1994; Freedman et al. 1994; Tanvir et al. 1995),
which indicate values of $H_0$ in the range 70 to 80 km sec$^{-1}$ Mpc$^{-1}$ . 
These figures have directed attention towards low-density models, in which the age problem might 
(just) be avoided. Similar considerations have been forced upon us by the discovery of a galaxy 
at redshift $z = 1.55$ that appears to be significantly older than the canonical universe at that 
redshift (Dunlop et al. 1996). These observations have also resurrected the cosmological constant 
as an option for serious consideration (Ostriker \& Steinhardt 1995), particularly low-density flat 
models($\Omega_0 + \Lambda_0 = 1$), which are compatible with inflationary models of the Universe
(Peebles 1984; Turner, Steigman \& Krauss 1984) and have ages which are significantly longer than $1/H_0$.

Here we report a study based upon one of the classical tests of observational cosmology, which is 
currently undergoing a renaissance (Kellermann 1993; Gurvits 1994; Jackson \& Dodgson 1996); 
we refer to the angular-size/redshift relation for ultracompact radio sources, with angular sizes 
determined by very long base-line interferometry (VLBI); our purpose is to decide which of the 
currently popular options (if any) is most favoured by the data, with particular reference to the age 
problem and the cosmological constant. Kellermann (1993) has argued that these sources should 
be free of the evolutionary and selection effects which have bedevilled attempts to use extended 
double radio sources in this context (Legg 1970; Miley 1971; Jackson 1973; Richter 1973; Masson 
1980; Barthel \& Miley 1988; Singal 1988; Nilsson et al. 1993; Daly 1994; Neeser et al. 1995), 
as they are deeply embedded in active galactic nuclei, and thus sheltered for example from the 
effects of an evolving extra­galactic medium; as a species they are at least as old as galaxies, but 
as individuals they have a fleeting existence, with lifetimes of perhaps 100 years, and it is thus 
reasonable to suppose that a stable population is established, characterized by parameters which 
do not change with cosmic epoch, particularly a mean linear size which is not a function of redshift. 
For further details and the associated caveats the reader is referred Kellermann's paper.

Kellermann presented a sample comprising 79 sources observed at a frequency of 5 GHz, divided 
into seven redshift bins. His main motivation was to show that the general form of the resulting 
angular-size/redshift diagram is cosmologically credible, and indeed compatible with a density 
close to the critical value $\Omega_0 = 1$, rather than to be specific about acceptable ranges for the
various cosmological parameters. Elsewhere (Jackson \& Dodgson 1996) we have looked at this sample in 
more detail, and produced a confidence region in three-dimensional parameter space comprising 
$\Omega_0$ and the cosmological constant $\Lambda_0$, and the mean projected linear size $d$ of
the population of compact sources. Here we propose to examine a much larger sample in similar detail.

\section{The Angular-Size/Redshift Relation for $z>0.5$}
 
Gurvits (1994) has presented a large VLBI compilation, based upon a 2.3 GHz survey due to 
Preston et al. (1985); the full sample comprises 337 sources with totally objective measures of 
angular size based upon fringe visibility (Thompson, Moran \& Swenson 1986, p. 13), and known 
redshifts. Their luminosities cover a wide range (almost 8 orders of magnitude), and are strongly 
correlated with redshift; this spread is greatly reduced (to less than 2 orders of magnitude) if sources 
with $z<0.5$ are ignored, and the corresponding redshift correlation is weak, so that the effects of 
a putative correlation between linear size and luminosity are minimised. In addition the reduced 
sample is morphologically homogeneous, in that it excludes radio galaxies, and comprises sources 
which are identified exclusively with quasars. Following Gurvits, we have thus considered a sample 
comprising 256 sources, the exact redshift range being 0.511 to 3.787, divided into 16 redshift bins. 
These are shown in Figure 1, in which each point represents the mean of $\log\theta$ for the 16 sources 
in each bin, and the error bars are $\pm$ the corresponding standard deviations divided by 4.
Each error bar is thus an estimate of the uncertainty in the mean, which according to our no-evolution 
hypothesis should be the same for each point, in reasonable accord with what we see.
 
As this sample was the subject of exhaustive statistical analysis in Gurvits' (1994) paper, we must 
stress what is new in our approach, and why we think that further analysis is justified. Gurvits 
considered four regression parameters, namely $\Omega_0$ and linear size $d$, plus a parameter representing 
a possible dependence of $d$ on radio luminosity, and finally one representing a possible dependence 
of $d$ on redshift, due to intrinsic cosmological evolution or to a dependence of linear size on emitted
frequency. The corresponding inverse problem (data to four-dimensional parameter space) is 
somewhat ill-conditioned, and these parameters were determined with rather large uncertainties. 
As Gurvits remarked, the addition of a fifth parameter to this set would hardly be worthwhile, and 
he quite explicitly excluded $\Lambda_0$ from his considerations. As this parameter is our primary interest 
here, we prefer to rely on Kellermann's plausibility arguments, which reduce our degrees of freedom 
to the three mentioned in the last paragraph of the Introduction, and allow a problem which is reasonably 
well-posed.

We have produced a number of one-parameter best-fitting curves, which assume that each point is taken 
from a normal distribution with standard deviation corresponding to the uncertainties discussed 
above; in each case the values of $\Omega_0$ and $\Lambda_0$ are fixed, and $\chi^2$ is minimised with respect
to $d$ as a free parameter. Figure 1 compares three such curves; the continuous one is the canonical CDM 
case, for which the optimum values are $d = 5.0$ pc and $\chi^2 = 26.8$; the dashed line corresponds to 
$\Omega_0 = 0.2$, $\Lambda_0 = -3.0$, for which the optimum values are $d = 4.8$ pc and $\chi^2 = 12.4$;
the dash-dotted curve is the best flat model $\Omega_0 = 0.2$, $\Lambda_0$ = 0.8, $d = 7.4$ pc, $\chi^2 = 16.4$.
 
The crucial qualitative feature is the pronounced minimum angular-size at $z = 1.25$ in the canonical 
CDM case (first discussed by Hoyle (1959) in the context of the steady-state cosmology), which 
is clearly at variance with the high-redshift end of the data set. Indeed the high-redshift points 
favour a curve which is asymptotically horizontal, which is a generic feature of low-density models 
(Jackson \& Dodgson 1996) with $\Lambda_0\leq 0$; this is why the dashed curve in Figure 1 is such a good 
fit, close to the 3-parameter global minimum $\chi^2 = 11.3$.  To further quantify these impressions, we 
have given free rein to $\Omega_0$, $\Lambda_0$, and $d$, to find global best-fitting parameters and the
corresponding 95\% confidence region, which is thus a volume in ($\Omega_0$, $\Lambda_0$, $d$) space,
$\chi^2\leq 22.4$.  Figure 2 is a view of this space from above, showing horizontal slices at the aforesaid
values $d = 4.8$ pc and $d = 7.4$ pc, and an intermediate value $d = 6.0$ pc. With regard to best-fitting values
of the three parameters our minimization routine typically produces values such as $\Omega_0 = 1.35$,
$\Lambda_0 = -63.0$, $d = 1.71$ pc, with $\chi^2 = 11.30$, but the problem is somewhat ill-posed in this respect;
there are many similar models, all with the same value of $d\sqrt{-\Lambda_0}$, for reasons which we have
explained in detail elsewhere (Jackson \& Dodgson 1996); such models would be ruled out by age considerations
alone.
 
Figure 3 is a projection of this three-dimensional region onto the $\Omega_0$--$\Lambda_0$ 
plane, in which the continuous curve is thus the envelope of curves similar to those shown in Figure 2,
and represents an overall 95\% confidence region which assumes no prior knowledge about the value of $d$;
the dashed curve is the corresponding 70\% confidence region. Figure 3 also includes the line
$\Omega_0 + \Lambda_0 = 1$, which defines spatially flat models, and a contour of constant age $t_0 = 0.56/H_0$, probably the lowest value which has ever been taken seriously ($t_0 = 14$ Gyr, $1/H_0 = 25$ Gyr).
We define the acceptable region to be the intersection of the 95\% confidence region with the 
regions $\Omega_0 + \Lambda_0\leq 1$, $H_0 t_0\geq 0.56$.

\section{Conclusions}

Concentrating first upon flat models, we see that the canonical inflationary CDM model (1,0) is 
quite unacceptable, being rejectable at the 98.5\% level of confidence; flat models with very low 
density are also unacceptable, $\chi^2$ rising rapidly to a value of 54.7 as the limiting case (0,1)
is approached. The most interesting feature here is that it is precisely those flat models with low to 
intermediate density, $\Omega_0\sim 0.1$ to $0.3$, (i.e. compatible with the internal dynamics of clusters of 
galaxies) which are favoured by these data. Nevertheless, none of these models are within the 70\% 
confidence region, which clearly favours models with $\Lambda_0\leq 0$; if we really believe that
$\Omega_0\sim 0.2$, then the best value compatible with our constraints is $\Lambda_0 = -3.0$.
Such models are open but violently decelerating, and destined to recollapse; we have argued elsewhere
(Jackson 1970; Jackson 1992; Jackson \& Dodgson 1996) that they are not in conflict with any of the classical cosmological tests, that is those which rely on observations out to $z\sim 4$. These conclusions are generally
in accord with those presented in our earlier paper (Jackson and Dodgson 1996), with reduced uncertainties 
as befits a larger and more homogeneous sample.
 
There are of course a number of caveats to be associated with the assumptions upon which these 
conclusions are based, which are discussed in detail by Kellermann (1993) and Gurvits (1994). 
Despite these reservations, we believe that this is a very promising approach to the determination 
of cosmological geometry. We have experimented with various rebinnings, which do not alter our 
conclusions: the data define a smooth curve, and despite the restricted z range, the shape of this 
curve is sufficiently well-defined to impose interesting constraints on the set of allowable models; 
its vanishing slope at high redshift is on the boundary between what is physically reasonable and 
what is not (Jackson \& Dodgson 1996), which if fictitious would require an improbable conspiracy 
of non-geometrical effects. In terms of the binary philosophy advocated in the first paragraph, the 
evidence presented here adds to the growing concensus (Ostriker \& Steinhardt 1995) that if the 
Universe is dominated by Cold Dark Matter, then it contains no more than the amount indicated 
the internal dynamics of clusters of galaxies.

\section{Acknowledgements}
Marina Dodgson acknowledges receipt of an UNN internal research studentship. We are very 
grateful to Leonid Gurvits, for sending us a listing of his VLBI compilation, and to Ken Kellermann
for several illuminating conversations.

\newpage
\References

\item[] Barthel P.D., Miley G.K., 1988, Nat, 333, 319 
\item[] Bolte M., Hogan C.J., 1995, Nat, 376, 399 
\item[] Daly R.A., 1994, ApJ, 426, 38 
\item[] Dunlop J., Peacock J., Spinrad H., Dey A., Jiminez R., Stern D., Windhorst R., 
\item[] 1996, Nat, 381, 581 
\item[] Freedman W.L. et al., 1994, Nat, 371, 757 
\item[] Gurvits L.I., 1994, ApJ, 425, 442 
\item[] Hoyle F., 1959, in Bracewell R.N., ed., IAU Symp. No. 9, Paris Symp. Radio Astronomy.
        Stanford Univ. Press, Stanford, p. 529 
\item[] Jackson J.C., 1970, MNRAS, 148, 249 
\item[] Jackson J.C., 1973, MNRAS, 162, 11P 
\item[] Jackson J.C., 1992, QJRAS, 33, 17 
\item[] Jackson J.C., Dodgson M., 1996, MNRAS, 278, 603 
\item[] Kellermann K.I., Nat, 361, 134 
\item[] Legg, T.H., 1970, Nat, 226, 65 
\item[] Masson C.R., 1980, ApJ, 242, 8 
\item[] Miley G.K., 1971, MNRAS, 152, 477 
\item[] Neeser M.J., Eales S.A., Law­Green J.D., Leahy P.J., Rawlings S., ApJ, 1995, 451, 76 
\item[] Nilsson K., Valtonen M.J., Kotilainen J., Jaakkola, T., 1993, ApJ, 413, 453 
\item[] Ostriker J.P., Steinhardt P.J., 1995, Nat, 377, 600 
\item[] Peebles, P.J.E., 1984, ApJ, 284, 439 
\item[] Pierce M.J., Welch D.L., McClure R.D., van den Bergh S., Racine R., Stetson P.B., 
        1994, Nat, 371, 385 
\item[] Preston R.A., Morabito D.D., Williams J.G., Faulkner J., Jauncy D.L., Nicolson G.D., 1985, 
        AJ, 90, 1599 
\item[] Richter G.M., 1973, Astrophys. Lett., 13, 63 
\item[] Singal A.K., 1988, MNRAS, 233, 87 
\item[] Tanvir N.R., Shanks T., Ferguson H.C., Robinson D.R.T., 1995, Nat, 377, 27 
\item[] Thompson A.R., Moran J.M., Swenson G.W., Jr, 1986, Interferometry and Synthesis in Radio 
        Astronomy, Wiley, New York 
\item[] Turner M.S., Steigman G., Krauss L.L., 1984, Phys. Rev. Lett., 52, 2090 

\endrefs

\newpage
\section{Figures}

\begin{figure}[here]
\begin{center}
\includegraphics[width=12.5cm] {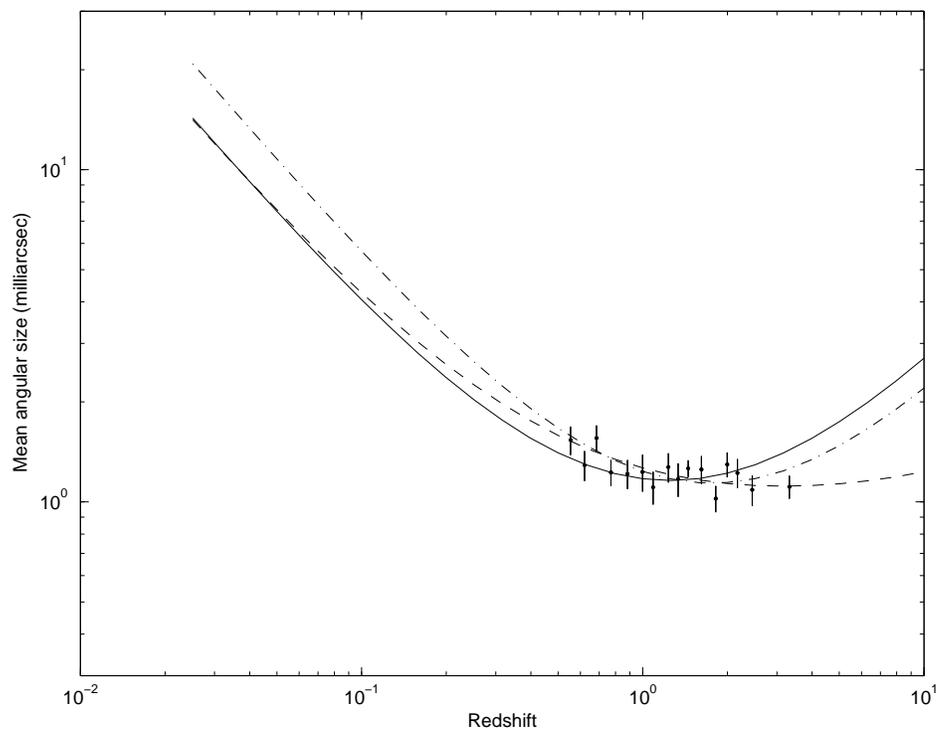}
\end{center}
\caption{\label{Fig1} Best-fitting angular-size/redshift curves for various universes, to be compared
with Gurvits' data points: the continuous line is the canonical CDM case $\Omega_0=1$, $\Lambda_0=0$;
the dashed line is a vacuum-dominated model with $\Omega_0=0.2$, $\Lambda_0=-3.0$;
the dash-dotted line the best flat model $\Omega_0=0.2$, $\Lambda_0=0.8$.} 
\end{figure}   

\begin{figure}[here]
\begin{center}
\includegraphics[width=12.5cm] {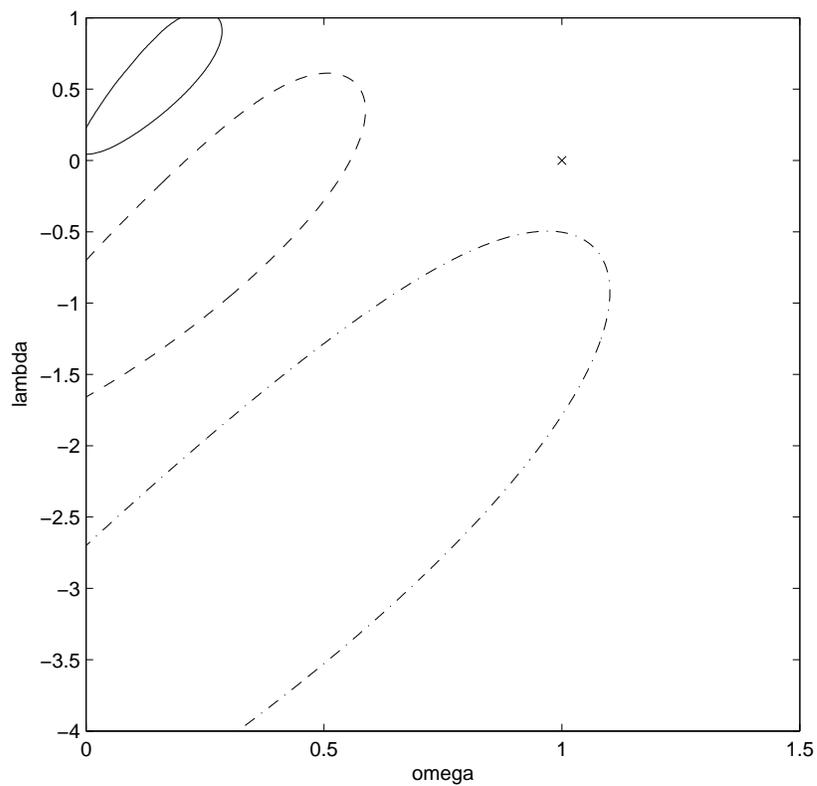}
\end{center}
\caption{\label{Fig2} Confidence regions (95\%) in the $\Omega_0$--$\Lambda_0$ plane, according to Gurvits data;
each contour corresponds to a horizontal slice through a three-dimensional region, with linear size $d$ as the vertical coordinate: the continuous line is for $d=7.4$ pc; the dashed line is for $d=6.0$ pc;
the dash-dotted line is for $d=4.8$ pc.} 
\end{figure}

\begin{figure}[here]
\begin{center}
\includegraphics[width=12.5cm] {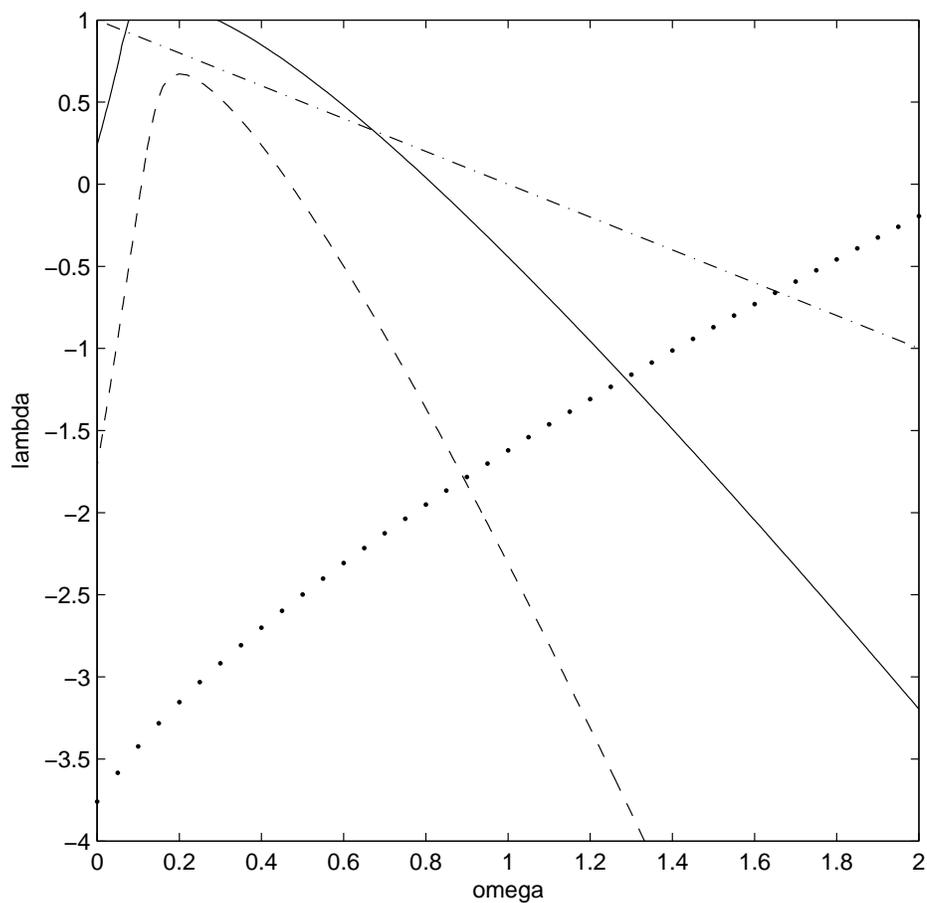}
\end{center}
\caption{\label{Fig3} Overall confidence regions in the $\Omega_0$--$\Lambda_0$ plane,
according to Gurvits data: the continuous line is for 95\%; the dashed line is for 70\%.
The dash-dotted line defines spatially flat models with $\Omega_0+\Lambda_0=1$.
The dotted line is a contour of constant age, $H_0t_0=0.56$.} 
\end{figure}

\end{document}